# Singularity and Similarity Detection from Signals Using Wavelet Transform

Hua-Liang Wei and S. A. Billings
Department of Automatic Control and Systems Engineering
The University of Sheffield
Mappin Street, Sheffield, S1 3JD
W.Hualiang@Sheffield.ac.uk, S.Billings@Sheffield.ac.uk

*Abstract*—The wavelet transform and related techniques are used to analyze singular and fractal signals. The normalized wavelet scalogram is introduced to detect singularities including jumps, cusps and other sharply changing points. The wavelet auto-covariance is applied to estimate the self-similarity exponent for statistical self-affine signals.

*Keywords-Brownian motion; fractals; jumps; sharp changes; self-similarity; singularies; wavelets.*

## I. Introduction

Sharp changes such as jumps, cusps and spikes are a commonly encountered singular phenomenon in practical signal processing. Another special class of singularities observed in real or artificial signals are fractals which are characterized by the property of self-similarity. These singularities and self-similarities can often be used to describe a wide class of practical phenomena arising in various fields of both science and engineering. The detection and analysis of these singularities and similarities have therefore attracted increasing interest in recent years, and many approaches including wavelet techniques have been proposed [3]-[6], [9], [12],[14].

It is well known that wavelets possess excellent local properties in both the time and the frequency domain, and can be used to approximate any square integrable signals, including signals with singularities. Let $f$ be a given signal defined in $L^2(R)$, the continuous wavelet transform (CWT) of the signal $f$ with respect to a given mother wavelet $\psi$ is defined as

$$W_f^\psi(b,a) = \frac{1}{\sqrt{a}} \int_{-\infty}^{\infty} f(t) \overline{\psi\left(\frac{t-b}{a}\right)} dt \quad (1)$$

where $a \in R^+$ and $b \in R$ are the scale (dilation) and translation (shift) parameters. The over-bar above the function $\psi$ indicates the complex conjugate.

The features of a given signal including the information of singularities carried on by the signal can often be reflected by the wavelet transform (1) and/or associated transformations. In this paper the wavelet scalogram will be introduced to detect singularities and similarities based on observed signals, and the wavelet auto-covariance will be used to estimate a self-similarity parameter of statistical self-affine signals.

## II. Singular Signals

Three classes of singular signals including sharp cusps, fractal geometry and statistical self-affine signals, are considered.

### A. Sharp Cusps and the Gibbs Phenomenon

Consider functions with an $\alpha$-power cusp [14]. A function $f$ is said to have an $\alpha$-power cusp at a point $x_0$ if there exists a positive constant $K$ such that

$$\lim_{h \to 0^+} |f(x_0 + h) - f(x_0)| \geq K |h|^\alpha \quad (2)$$

Clearly, if $\alpha = 0$, then $f$ has a jump at $x_0$. In this study, the parameter $\alpha$ will be restricted to the interval $0 \leq \alpha < 1$.

Although most discontinuous signals can be recovered from the inverse wavelet transform with good convergence rates, it has been shown [9],[10] that a wide class of wavelet expansions suffer from Gibbs phenomenon, which states that the pointwise convergence of global approximation of discontinuous functions is at most first order. In the presence of jumps global approximations are often oscillatory. Gibbs phenomenon for a given discontinuous signal can be reflected by the wavelet transform of the signal due to the property that big jumps often produce large wavelet coefficients.

In this study the wavelet scalogram(WS) will be used to detect sharp cusps. The wavelet scalogram of a signal $f$ is related to the associated wavelet transform by

$$S_f^\psi(b,a) = \frac{|W_f^\psi(b,a)|^2}{a} \quad (3)$$

### B. Fractal Geometry

The term *fractal*, introduced by Mandelbrot [7], has now been extended and developed into three related concepts: *geometric*, *temporal* and *statistical* fractals. Generally, the concept of a fractal is defined in terms of self-similarity. Many initial applications of fractals focused on problems in condensed matter and computer graphics. Soon afterwards, the applications have been extended to aid the understanding of a great number of natural phenomena which exhibit self-similarity in biology, ecology, geometry, engineering, music, physics, physiology and topography.

Various approaches have been proposed to produce fractal models. The concept of an iterated function system (IFS) [1] provides an important tool to generate fractal images. In particular, an IFS with probabilities is defined as the following structure:

$$\Sigma: \quad <(X,d),W> \quad (4)$$

where $(X,d)$ is a complete metric space with a metric $d$; $W$ is a family of a finite set of contraction mappings $\{w_m : m = 1,2,\cdots,M\}$, and each $w_m$ is defined on the space $(X,d)$; $P$ is an ordered set of numbers $\left\{p_m : \sum_{m=1}^{M} p_m = 1, p_m > 0, m = 1,2,\cdots,M\right\}$, and a probability $p_i$ is to be assigned to the $i$-th mapping $w_i$. Choose $\mathbf{x}_0 \in X$, and then produce $\mathbf{x}_n (n \geq 1)$ recursively and independently as

$$\mathbf{x}_n \in \{w_1(\mathbf{x}_{n-1}), w_2(\mathbf{x}_{n-1}), \cdots, w_M(\mathbf{x}_{n-1})\} \quad (5)$$

where the probability of the event $\mathbf{x}_n \in w_m(\mathbf{x}_{n-1})$ is $p_m$.

As a special case of an IFS with probability, the affine model below can be used to produce very rich 2-D fractal images

$$w(\mathbf{x}) = w\begin{pmatrix} x_1 \\ x_2 \end{pmatrix} = \begin{pmatrix} a & b \\ c & d \end{pmatrix}\begin{pmatrix} x_1 \\ x_2 \end{pmatrix} + \begin{pmatrix} e \\ f \end{pmatrix} \quad (6)$$

where $a$, $b$, $c$, $d$, $e$ and $f$ are randomly set to some specified real values.

Table 1 gives a 2-D fractal model, which can produce very complex patterns, e.g., a group of 'trees' as shown in Fig. 1. As will be shown, the self-similar feature of fractal signals produced by the affine model (6) can be detected by the wavelet transform.

TABLE I. COEFFICIENTS AND PROBABILITIES OF AN AFFINE FRACTAL MODEL

| Probabilities | Coefficients | | | | | |
|---|---|---|---|---|---|---|
| | $a$ | $b$ | $c$ | $d$ | $e$ | $f$ |
| $p_1 = 0.01$ | 0 | 0 | 0 | 0.16 | 0 | -1.00 |
| $p_2 = 0.85$ | -0.85 | -0.04 | -0.04 | 0.85 | 0 | 1.60 |
| $p_3 = 0.07$ | -0.20 | 0.26 | 0.23 | 0.22 | 0 | 1.60 |
| $p_4 = 0.07$ | 0.15 | -0.28 | 0.26 | 0.24 | 0 | 0.22 |

C. *Statistical Self-Similar Signals*

A self-affine set is statistically invariant under an affine transformation. A 2-dimensional surface described by a function $f(x,y)$ is a self-affine fractal, if there exists a number $H$ such that

$$f(x,y) = \lambda^{-H} f(\lambda x, \lambda^H y) \quad (7)$$

where $\lambda$ is a positive number, $H$ is called the *Hurst* exponent or *Hausdorff* exponent or self-affine exponent with a value between 0 and 1, that is, $0 \leq H \leq 1$. Equation (7) indicates that $f(\lambda x, \lambda^H y)$ is statistically similar to $f(x,y)$ with a similarity exponent $H$. In one-dimension, a self-affine fractal [6],[13] is defined as $f(x) = \lambda^{-H} f(\lambda x)$. In this case, $x$ and $f(x)$ are often interpreted as the time and the corresponding trajectory (position), respectively. It has been proved [13] that the *Hurst* exponent, $H$, and the self-affine fractal dimension, or the box-counting dimension, $D$, are related by the equation $H = 2-D$. Therefore $1 \leq D \leq 2$ corresponds to $0 \leq H \leq 1$ for a self-affine fractal. If $H=1$, the self-affine fractal becomes self-similar, which is by definition isotropic. A stochastic process or a surface with $H > 1/2$ is said to be persistent, and that with $H<1/2$ is said to be antipersistent.

Following [6], the basic property of a self-affine time series is that the power spectral density of the time series has a power-law dependence on frequency. The physical features of a dynamic system can be easily detected and revealed using frequency domain analysis, which is often implemented by means of Fourier transforms of the covariance functions. One feature of a self-affine time series is that the power spectral density of the time series has a power-law dependence on frequency

$$P(\omega) \propto |\omega|^{-\beta} \quad (8)$$

Note that for a relationship between the power-law exponent $\beta$, the Hurst exponent $H$, and the fractal dimension $D$ is given [13] by $\beta = 2H + 1 = 5 - 2D$. For a self-affine process [6], $0 \leq H \leq 1$, $1 \leq D \leq 2$ and $1 < \beta < 3$. For a Brownian motion, $H = 0.5$, $D = 1.5$ and $\beta = 2$.

Following [3], the covariance of the wavelet transform (1) of a random signal $x(t)$ at a given scale $a$ can be defined as

$$R_x^\psi(t,s;a) = E[W_x^\psi(t,a)\overline{W_x^\psi(s,a)}] \quad (9)$$

It has been shown [15] that in the case that $x(t)$ is a self-affine signal obeying the power-law (8), the *auto-covariance* of the wavelet transform of the signal $x(t)$ also obeys a power-law in the sense that

$$R_x^\psi(a) = E[W_x^\psi(t,a)\overline{W_x^\psi(t,a)}]$$

$$= \frac{a}{2\pi} \int_{-\infty}^{\infty} \frac{|\hat{\psi}(a\omega)|^2}{|\omega|^\beta} d\omega$$

$$= \frac{a^\beta}{2\pi} \int_{-\infty}^{\infty} \frac{|\hat{\psi}(\omega)|^2}{|\omega|^\beta} d\omega = C a^\beta \quad (10)$$

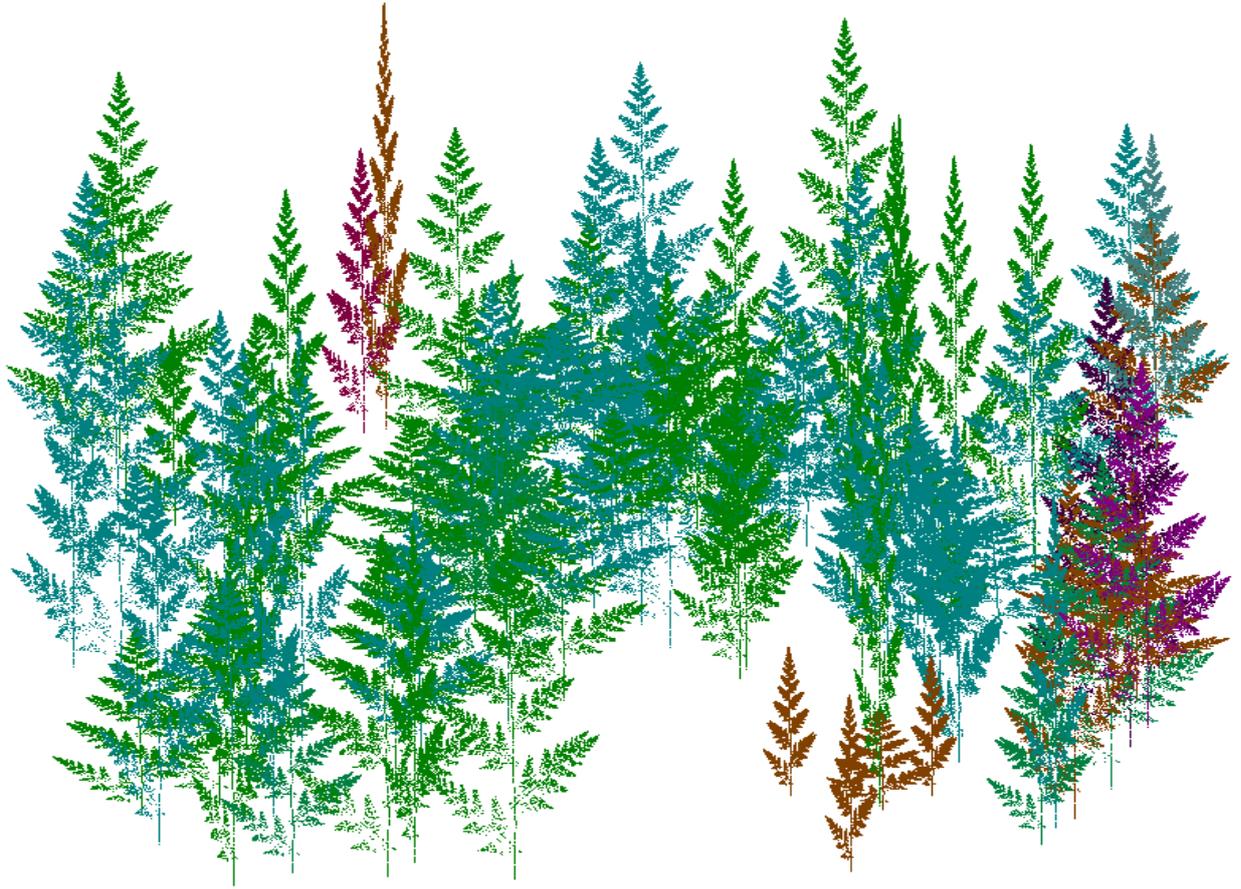

Figure 1  A group of trees produced from the model (6) with the coefficients listed in Table (1)

where $C = \frac{1}{2\pi}\int_{-\infty}^{\infty}|\hat{\psi}(\omega)|^2 / |\omega|^\beta \, d\omega$ is a constant. This suggests that for a self-affine process $x(t)$, the auto-covariance of the wavelet transform of signal $x(t)$ also obeys the power-law with respect to the wavelet scale parameter $a$. Therefore, the new result (10) can be used to estimate the power-law exponent of self-affine fractals.

### III. Numerical Examples

This section provides several examples to illustrate the applications of wavelet transformations for detecting and analyzing singular and self-similar signals.

#### A. Jump and Cusp Detection

Two signals, which contain both jumps and cusps and which were contaminated by noise were considered to demonstrate the application of the wavelet scalogram defined by (3).

*Example 1—a chirp signal with a jump and a sharp cusp*

The original noise free signal is show in Fig. 2(a), where there is a jump point at 0.5 and a cusp point as 0.6; the noise corrupted signal is show in Fig. 2(b), where the added noise was a Gaussian white signal with zero mean and standard derivative $\sigma = 2$. The 3-D wavelet scalogram for this signal is shown in Fig. 3, where both the jump and the cusp points were precisely detected.

*Example 2—a signal with a jump, a sharp cusp and some smooth bumps*

The model of the signal is given by [13]

$$f(x) = 2\sin(4\pi x) - 6|x - 0.4|^{3/10} \\ - 0.5 sign(0.7 - x) + \varepsilon \qquad (11)$$

where $\varepsilon$ was a Gaussian white noise with zero mean and standard derivative $\sigma = 0.2$. The noise free and the noise corrupted signals are shown in Fig. 4(a) and (b).

The 3-D wavelet scalogram for this signal is shown in Fig. 5. As expected, both the jump and the cusp points were accurately detected.

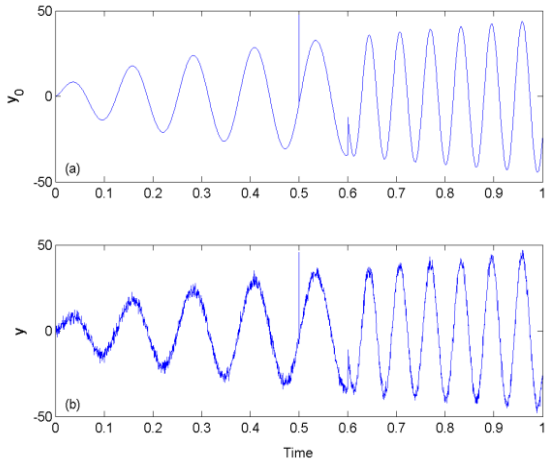

Figure 2 The signal described in Example 1.
(a) noise free; (b) with a noise.

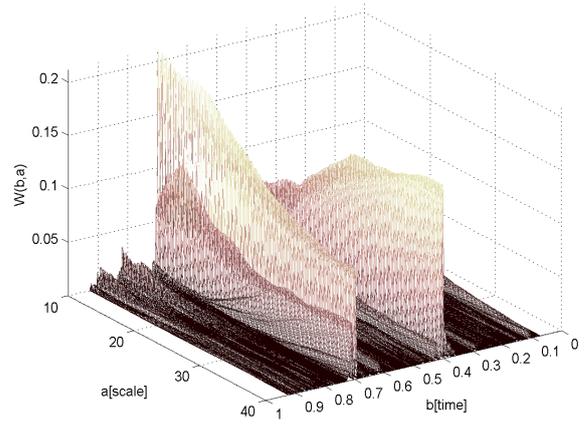

Figure 5 The wavelet scalogram for the signal described by
(11) in Example 2.

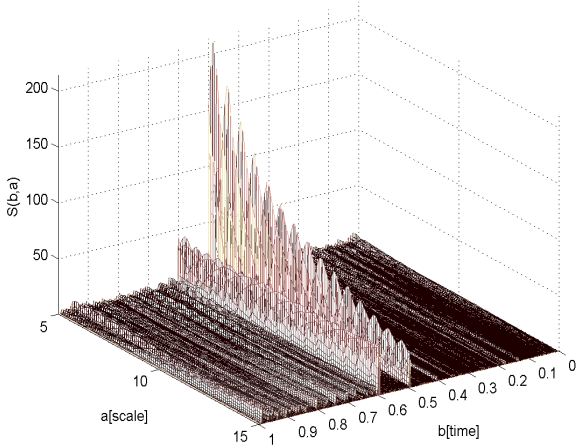

Figure 3 The wavelet scalogram for the signal described
in Example 1.

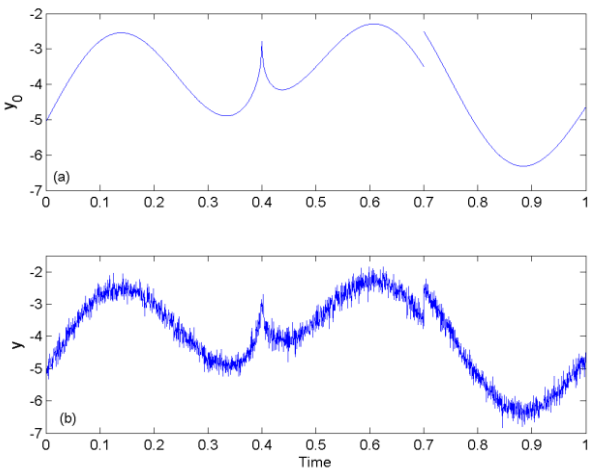

Figure 4 The signal described by (11) in Example 2.
(a) noise free; (b) with a noise.

## B. Self-Similarity Detection for Fractal Geometry

The self-similarity hidden in fractal geometry can be detected by the wavelet transform.

### Example 3—a fractal tree

Take the fractal Barnsley tree as an example. With the coefficients listed in Table 1, two scalar signals $x_1(k)$ and $x_2(k)$ were generated using the affine transform (6), and part of these are shown in Fig. 6 (a) and (b). The two signals $x_1(k)$ and $x_2(k)$ can form a Barnsley tree similar to those shown in Fig. 1. As will be seen, the self-similarity hidden in the signals can be detected by the wavelet transform. The maxima of the wavelet transform of the two signals are shown in Fig. 7, from which the self-similarity possessed by the signals can be clearly seen.

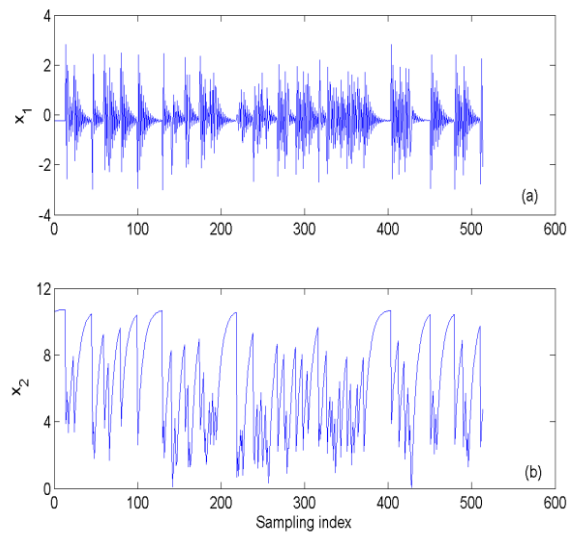

Figure 6 The signals generated from the model
given by Table 1.

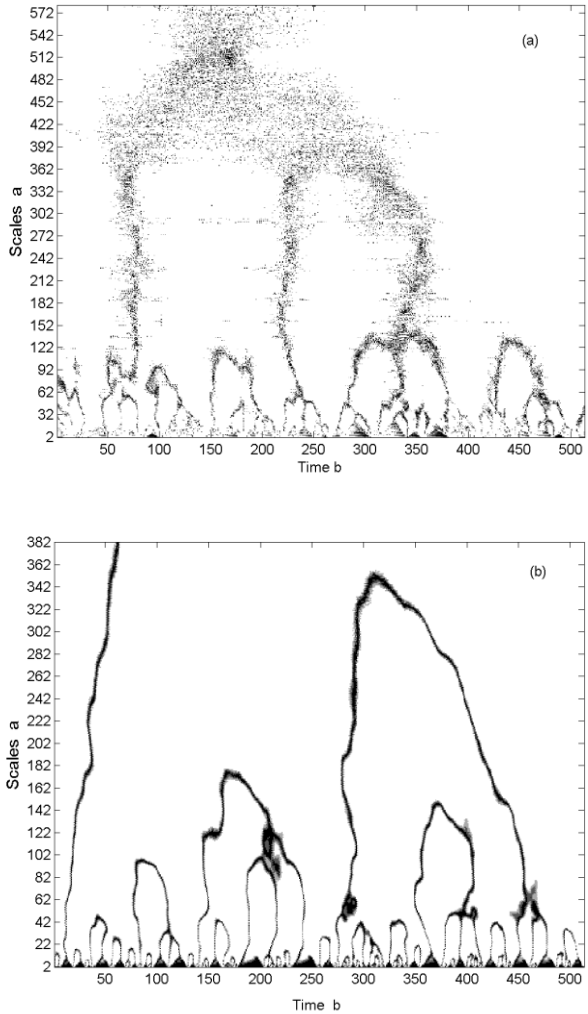

Figure 7 The wavelet transform modulus maxima for the Barnsley tree signals shown in figure 6. (a) for $x_1$; (b) for $x_2$.

## C. Estimation of a Self-Similar Parameter

The wavelet transform provides an effective tool to estimate the self-similar parameter for self-affine stochastic processes, where the power spectral density of the time series has a power-law dependence on frequency.

*Example 4—a simulated fractional Brownian motion*

Fig. 8(a) shows one realization of a fraction Brownian motion with a Hurst exponent $H$=0.8. From Fig. 8(b), the slope of the line was calculated to be $\beta \approx 2.658$, and the Hurst exponent for the simulated Brownian motion was estimated to be $\tilde{H} \approx (\beta-1)/2 \approx 0.829$.

*Example 5—a real data set*

Fig. 9(a) shows a set of sales data taking from [2]. This is a nonstationary time series. The slope of the wavelet auto-covariance line was approximately $\beta \approx 1.97$, the Hurst exponent for this process was $H \approx 0.485$, which indicates that the nonstationary time series can be viewed as a Brownian motion.

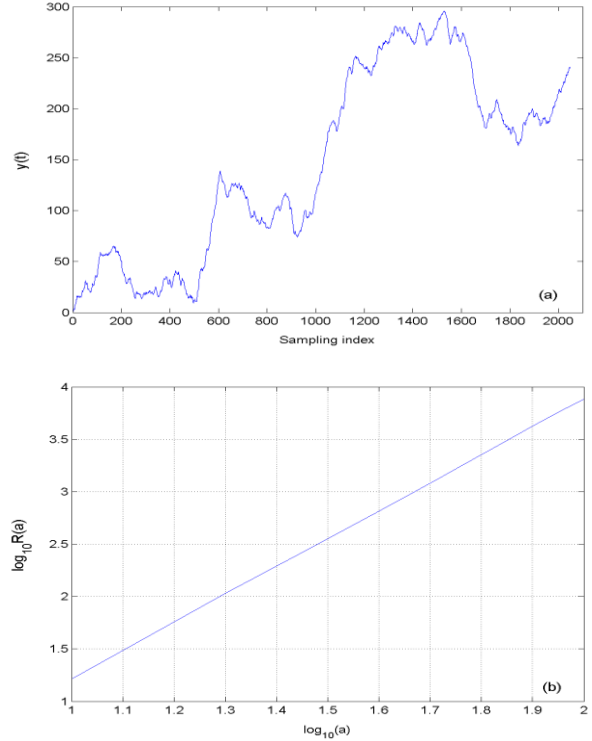

Figure 8 One realization of a Brownian motion and the associated wavelet covariance. (a) a Brownian motion; (b) the wavelet covariance defined by (10).

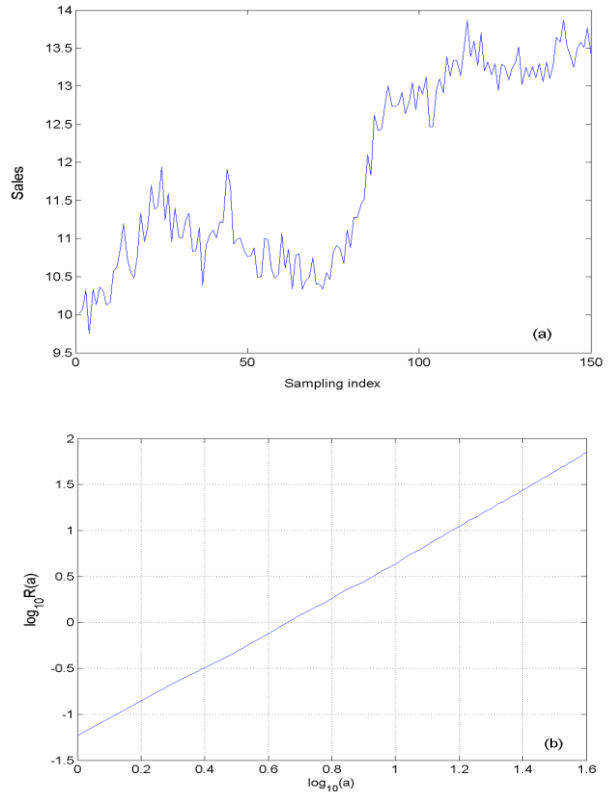

Figure 9 The sales data described in Example 5 and the associated wavelet covariance. (a) sales data; (b) the wavelet covariance defined by (10).

## IV. CONCLUSIONS

The wavelet transform and related techniques provide powerful tools for detecting and analyzing singular and self-similar signals. Information carried by a singular signal will be totally preserved in the wavelet coefficients, and sharply changing points can be sensitively detected using the wavelet scalogram, which is a variant of the normal wavelet transform. The self-similarity that exits in fractal geometry can also be revealed by the wavelet transform. It has been shown that the auto-covariance of the wavelet transform of a self-affine signal obeys a power-law. This result enables the wavelet transform to be used to estimate the self-similar exponent for self-affine stochastic processes.

## ACKNOWLEDGMENT

The authors gratefully acknowledge that this work was supported by EPSRC (UK).

## REFERENCES

[1] M. B. Barnsley, *Fractals Everywhere*. Boston: Academic Press, 1988.

[2] G. E. P. Box, G. M. Jenkins, and G. C. Reinsel, *Time Series Analysis: Forecasting and Control* (3rd ed.). New Jersey: Prentice Hall, 1994.

[3] P. Flandrin, "On the spectrum of fractional Brownian motions," IEEE Trans. Inform. Theory, vol.35, pp. 197-199, January 1989.

[4] L. Hongre, P. Sailhac, M. Alexandrescu, and J. Dubois, "Nonlinear and multifractal approaches of the geomagnetic field," Physics of the Earth and Planetary Interiors, vol. 110, pp. 157-190, Feb. 1997.

[5] B. D. Malamud and D. L. Turcotte, "Self-affine time series: measures of weak and strong persistence," Journal of Statistical Planning and Inference, vol. 80, pp. 173-196, Aug. 1999.

[6] S. Mallat and W. L. Hwang, "Singularity detection and processing with wavelets," IEEE Trans. Inform. Theory, vol.38, pp. 617-643, March 1992.

[7] B. B. Mandelbrot, *The Fractal Geometry of Nature*. New York: W.H.Freeman and Company, 1983.

[8] Mandelbrot and J. W. Van Ness, "*Frational Brownian motion, fractional moise and application*," SIAM Review, vol.10, pp. 422-437, October 1968.

[9] J. F. Muzy, E. Bacry, and A. Arneodo, "Wavelets and multifractal formalism for signals: application to turbulence data," Physical Review Letters, vol. 67, pp. 3515-3518, Dec. 1991.

[10] D. K. Ruch and P. J. Van Fleet, "Gibbs' phenomenon for non-negtive comapctly supported scaling vectors," J. math. Anal. Appl., vol. 304, pp. 370-382, April 2005.

[11] H. T. Shim and H. Volkmer, "On the Gibbs phenomenon for wavelet expansions," J. Appr. Theory, vol. 84, pp. 74-95, Jan. 1996.

[12] S. Soltani, P. Simard, and D. Boichu, " Estimation of the self-similarity parameter using the wavelet transform," Signal Processing, vol. 84, pp. 117-123, Jan. 2004.

[13] R. F. Voss, "*Fractals in nature : from characterization to simulation*," in The Sceince of Fractals Images, H.O. Peitgen and D. Saupe (Eds.), Springer, New York, 1988.

[14] Y. Z. Wang, "Jump and sharp cusp detection by wavelets," Biometrika, vol. 82, pp. 385-397, June 1995.

[15] H. L. Wei, S. A. Billings, and M. Balikhin, "Analysis of the gemagnetic activity of the Dst index and self-affine fractals using wavelet transforms," Nonlinear Processing in Geophysics, vol. 11, pp. 303-312, June 2004.